\documentstyle[prl,preprint,aps,psfig]{revtex}

\begin{document}
\title{DYNAMICAL TRANSITIONS IN CORRELATED DRIVEN DIFFUSION IN A 
PERIODIC POTENTIAL.}
\author{Oleg M. BRAUN, Thierry DAUXOIS, Maxim V. PALIY, and Michel
PEYRARD}
\address{Laboratoire de Physique, URA-CNRS 1325,\\
ENS Lyon, 46 All\'{e}e d'Italie, 69364 Lyon C\'{e}dex 07, France}
\date{\today}
\maketitle

\begin{abstract}
The diffusion of a two-dimensional array of particles driven by a
constant force in the presence of a periodic external potential
exhibits a hierarchy of dynamical phase transitions when the driving
force is varied. This behavior can be explained by a simple
phenomenological approach which reduces the system of strongly
interacting particles to weakly interacting quasi-particles
(kinks). The richness of the strongly coupled system is however not
lost because, contrary to a single-Brownian particle, the array shows
an hysteretic behavior even at non-zero temperature. The present
investigation can be viewed as a first step toward understanding
nanotribology. 

\end{abstract}


The diffusion of a Brownian particle, driven by an external force and
subjected to a periodic potential, is a situation which arises in
several fields of science\cite{Risken1984} such as solid-state
physics, surface physics\cite {Villainpimpinelli}, chemical physics
and even communication theory. It is now well understood and provides
a simple example of an out-of-equilibrium phase transition, between a
locked and a running state. The case of many interacting particles is
even more interesting because collective effects modify qualitatively
the picture giving rise, in some parameter range, to a dynamical state
which is very reminiscent of {\em a traffic jam at the atomic scale.}
We discuss here this behavior in the context of solid state friction
because it provides a typical example which may be amenable to
experimental tests, but the basic ingredients required to observe the
phenomenon are simple and can be found in many physical systems. 

Understanding the diffusion and mobility of strongly interacting atoms
subjected to a periodic potential and
driven by an external force is a first step
toward understanding solid friction at the atomic level\cite{tribology}.
For noninteracting  atoms the problem would be simple since 
it essentially reduces to the
diffusion of a Brownian particle in a periodic potential.
Under the influence of a dc force $F$ it will preferably
diffuse in the direction of the force and in average there will be a
drift velocity $\langle v\rangle $ which depends on $F$. The mobility
$B$ is then defined as $B=\langle v\rangle /F$.  If, for small forces,
$B$ is independent on $F$ (the linear response regime), for arbitrary
ones a nonlinear response takes place, and the task is to calculate
this nonlinear mobility.

The total potential experienced by the Brownian particle is the sum of
the periodic potential and the potential $-Fx$ due to the driving
force, i.e.\ it corresponds to a corrugated plane, with an average slope
is determined by $F$.  At small forces the potential has local minima,
therefore the particle is static and its mobility vanishes. On the
contrary for large forces there are no stable positions, and the
particle slides over the corrugated potential, reaching its maximum
mobility $B_f=(m\eta )^{-1}$, where $m$ is the mass of the particle
and $\eta $ is the friction coefficient. A simple calculation shows
that, denoting by $\varepsilon $ the height of the periodic potential,
$a$ the lattice constant, and $c$ a constant depending on the shape of
potential, the critical force for which the stable positions disappear
is $ F_r^0=c\,{\varepsilon /a}$.

However, for the underdamped case, the system may also have a running
solution, even if the minima of the potential do exist. Indeed,
because of their momentum the particles may overcome the next hill,
which is lower than the one from which they are falling due to the $-Fx$
contribution to the potential, if the gain in potential energy is
greater than the energy dissipated during this motion.  One finds that
this second critical force is $F_\eta ^0=c^{\prime }\eta
\sqrt{m\varepsilon }$, where $c^{\prime }$ is a constant which depends
on the potential shape. As the particle is either locked or running,
depending on its initial velocity, the system exhibits bistability and
the transition between these two states shows hysteresis. However,
{\em for a single particle the bistability disappears in the presence
of an external noise such as thermal fluctuations, no matter how small
the noise is,} because the fluctuations can kick the particle out of the
locked state by usual thermal activation.  Thus, the Brownian motion
of a single particle driven by an external force shows hysteresis
only for zero temperature.

\bigskip

The case of interacting particles in a periodic potential is a much
more difficult problem; interesting numerical results were obtained in
for high damping, when the time-independent Schmoluchowsky
equations may be reduced to a one-particle equation with an effective
on-site potential and then solved numerically by the transfer-integral
method~\cite {ButtikerLandauer1981}. The results show that there is a
region of highly nonlinear mobility but without a bistability
phenomenon. Besides, recently Persson~\cite{persson} has used
molecular dynamics to study a 2D system of interacting atoms subjected
to a periodic potential. In the underdamped case, he observed a
dynamical phase transition similar to the one-particle case. Recalling
the well-known Aubry transition from the pinned state to the freely
moving state in the Frenkel-Kontorova model with an incommensurate
atomic concentration~\cite{Aubry}, Persson supposed that his results
could be explained in a similar way.

In the present Letter, we study the driven motion of interacting atoms
in a 2D potential for arbitrary damping.  First we show that the final
state observed by Persson corresponds in fact to the sliding motion of
atoms in an inclined potential. Then we demonstrate that the
transition from the locked to the sliding state passes through a
hierarchy of hysteretic depinning transitions. An important point is
that  {\em hysteresis persists in the presence of thermal fluctuations.}
Finally, we analytically compute the critical forces and the
corresponding mobilities using a phenomenological
approach~\cite{bklast} which treats a system of strongly interacting
atoms as a system of weakly interacting quasiparticles (kinks).

Although it is still oversimplified, the generalized Frenkel-Kontorova
model that we consider provides a rather complete description of
a layer of atoms adsorbed on a 2D crystalline surface when the motions in
the vertical dimension are also taken into account. The parameters are
chosen for the adsystem Na-W(112) and this model was
proposed~\cite{theta} to explain the intriguing experimental results
obtained for the dependence of the diffusion coefficient of strongly
interacting adatoms when the concentration varies.

The model considers the following Langevin equation for the
$x$-coordinate of $i$-th atom, 
\begin{eqnarray}
&m&\ddot{x_i} + m \eta \dot{x_i} + \frac{d}{d x_i} \biggl[
V_x(x_i)+V_y(y_i) + \frac{1}{2} m \omega_z^2 z_i^2
\nonumber \\
&+&\sum_{j\neq i}
V_0\exp (-\beta _0|\vec{r}_i-\vec{r}_j|)
\biggr]=F^{(x)} + \delta\!F_{i}^{(x)} (t),
\end{eqnarray}
and similar equations for the coordinates $y_i$ and $z_i$. The particles
are therefore in a periodic rectangular potential,
\begin{equation}
\label{pr}V_{\alpha}(\alpha) = \frac{\varepsilon_{\alpha}}{2} \frac {
(1+s_{\alpha})^2[1-\cos (2\pi \alpha/a_{\alpha})]}
{1+s_{\alpha}^2-2s_{\alpha}\cos (2\pi \alpha/a_{\alpha})},
\end{equation}
where $\alpha$ is $x$ or $y$, and  parameters
$s_x=0.2$, $s_y=0.4$, $\varepsilon_x=0.46$ eV, $\varepsilon_y=0.76$ eV, 
$a_x=2.74\, \AA$, $a_y=4.47\, \AA$ are chosen in relation to the highly 
anisotropic {\em channeled} surface W(112) \cite{theta}. Atomic mass is
$m=1$, 
and we put $\omega_z = 1.84$, $\eta=0.165$ in the corresponding units. 
The exponential interaction law with $V_0=10$ eV, $\beta_0=0.85\,
\AA^{-1}$ 
corresponds to the repulsion of the adatoms at rather high
concentration,
$F^{(x)}$ is the external force applied {\em along the channels}, 
and $\delta\!F$ is a gaussian random force simulating the interaction
with a 
thermal bath. We start with a ground state of the system, then the
temperature 
and later the force are increased adiabatically. 
We then compute the mobility for different values of $F$ and $T$.

An important parameter is the atomic concentration $\theta $,
corresponding to the ratio between the number of particles $N$ and the
number of available sites $M$. In this Letter we present data at
several generic concentrations in order to show various aspects,
although the general statements are common for the wide range of
atomic concentrations.

Let us first consider the case of $\theta =21/41$ modeled by 30
channels, each with $N=105$ and $M=205$. As this concentration, close
to $\theta = 1/2$, is not a
simple $1/q$ value with $q$ integer, the ground state
corresponds
to large regions where the effective concentration is $1/2$,
equidistantly
separated by zones of compression~\cite{theta}. Since these zones in the
standard FK model are called kinks (or antikinks in the case of
localized
expansions), we use this terminology here too. The number of kinks
 is proportional to the difference between $\theta $ and the
closest
value $1/q$~\cite{bklast}. All the parameters, such as the mass of kink,
Peierls-Nabarro potential, and energy of creation of kink-antikink
pair can be easily evaluated~\cite{theta}. In spite of the obvious
difference with the mathematical description of soliton behavior in
nonlinear partial differential equations, we will show that there are
surprising similarities and moreover that they allow us to explain
{\em qualitatively} and {\em quantitatively} the
intriguing behavior of the system in a simple way.

A generic evolution of the mobility as a function of the external
force is shown on Fig.~\ref{fig1}. One first notice a hysteresis and
distinguish four different regions when the external force is
increased. In the very low force range, the mobility is zero. Above a
critical force $F_k$, the mobility jumps to a first plateau at $B_k$. 
Then there is a second
plateau around $B_m$ for $ F>F_{pair}$, and finally for a force
higher than $F_r$, the system reaches the maximum mobility
$B_f$. When the force is decreased, the systems jumps directly
to the static state for forces lower than $F_\eta $. 
Let us explain these different states.

In the first region, the force is too low, and neither the kinks nor
the individual atoms move. Both are trapped in their wells and the
system is in the {\em locked state}. When the force reaches a critical
value $F_k$, the mobility jumps to a nonzero value $B_k$.  This state
corresponds to the {\em kink-running state}. Indeed, a careful study
of the time dependence of positions of all atoms indicates clearly
that the compressed zones are moving but not the individual atoms, except
when a kink passes through their site. A simple analogy with a
single particle in the periodic potential allows to compute this
critical force. Owing to the lattice discreteness, a kink in the FK
model moves in the periodic Peierls-Nabarro potential, whose barrier
$E_k$ could be easily determined~\cite{theta}; in addition, the
potential is tilted due to the external force. This quasiparticle
(kink) will be trapped until the last stable state disappears, i.e.\
when $F_k=c{ E_k / a}$. Besides, as we know the number of kinks
$\theta_k$ in the system, the mobility of the kink-running state is
$B_k=\theta_k B_f$. This approach is very successful as shown in
Fig.~\ref{fig1}, where the values $B_k$ and $F_k$ are shown.

In the kink-running domain, the atoms are static contrary to the kinks
since the energy barrier $\varepsilon$ is always greater than the
Peierls-Nabarro barrier $E_k$. Physically this phenomenon signifies
that it is easier to move a dislocation coherently than to move all
the atoms. In addition, a detailed study shows a decrease of the
relative distance between the kinks. One finds a tendency for the kinks
to bunch like the cyclists in a ``peloton''.  The probable reason is
that, as the kinks are not exact solutions of the system and therefore
radiate waves, the oscillatory tail of a kink could help the following
kink to overcome the Peierls barrier and
to catch up with the previous one. A second point to notice, is that this
motion of kinks shows also a hysteresis even for non zero temperature.
 If one decreases the force
when this first plateau is reached, the system goes back to the locked
state for a critical force lower than $F_k$. This indicates that
kinks have a behavior more complex than a single Brownian particle
although the plateau at $B_k$ is well predicted by the one-particle
picture. 

In the range of force corresponding to the second plateau, at
$F>F_{pair}$, additional kinks and antikinks could be created since
the energy barrier for the nucleation of new pairs
vanishes. Therefore, we have to take into account not only the
geometrical (ground state) kinks but also the ``force-excited''
kinks~\cite{bklast}. Fig.~\ref{fig2} presents the time dependence of
the positions of all atoms. Two different regions can be seen, mobile
and immobile ones. The finite time interval between the snapshots
results in a stroboscopic effect giving a wrong impression for atomic
trajectories; in order to show one actual trajectory we marked one of
atoms by black diamonds while others are indicated by unfilled
diamonds. The picture is very reminiscent of a traffic jam: the
particle is trapped into an immobile zone, until being first in this
region; then the particle ballistically move till the next
high-density zone, where it is stopped again. The velocity probability
presents a two-bells shape, corresponding to static and moving atoms:
this is a {\em coexistence regime}.

A careful examination of the $B(F)$ dependence in this range of $F$ 
shows that the mobility
is even slightly decreasing with increasing force. Indeed, after the
kink-antikink nucleation threshold $F_{pair}$, the kinks start to
bunch into compact groups, as in the context of Josephson
junction~\cite{ustinov}.  A simple phenomenological theory gives
$B_m(\theta) \propto B_f\, \,(1 -\theta)/\theta$,
which agrees very accurately with the numerical
results~\cite{pourpre}. Moreover, the study shows that the mobile
subsystem corresponds to a bunch of the antikinks whereas the immobile
one corresponds to a bunch of the kinks.

Finally, after the coexistence regime, for high enough force, all
atoms are sliding over the periodic potential and the mobility reaches
its maximum value $B_f$. When the force is reduced, the description of
the dynamics in terms of kinks has lost its meaning since there is no
special organization that subsists.  This is why the system goes back
directly to the locked state at the critical value $F_\eta$.

Thus, the system of strongly interacting particles in an anisotropic
external potential does present a dynamical phase transition when the
dc external force is varied. We are able to explain the multiple steps
by the {\em hierarchy of depinning}: first, the geometrical kinks,
then the ``force excited'' kinks and finally the atoms. But it is more
remarkable that this behavior does not disappear with thermal
fluctuations as the hysteresis of a single driven particle in a
periodic potential. Fig.~\ref{fig3} attests that the hysteresis
survives for finite temperatures.

Let us derive an approximate expression for the critical force
$F_\eta$ versus temperature. At zero temperature the
back transition approximately corresponds for low concentration to the
external single-particle threshold $F_\eta^0$. 
For nonzero temperatures, the system gets
locked when the probability for the velocities to be
lower than $F_\eta^0/m\eta$ is greater than a threshold~$P_c$. As
a result, the critical force follows the law
\begin{equation}
F_\eta=F_\eta^0+\sqrt{2mk_BT\eta^2}\ {\rm erf}^{-1}(1-2P_c)
=F_\eta^0+\delta
\sqrt{T},
\end{equation}
where ${\rm erf}^{-1}$ is the inverse of the error function.  With
$F_\eta^0\simeq0.144$, the solid curve in Fig.~\ref{fig3} shows that
this expression scales very accurately with above expression if
$\delta=0.35$.

Unfortunately, we do not have a complete understanding of evolution of
the forward transition versus temperature. Because of temperature
fluctuations, the particles  feel a smoother potential and a
smaller barrier to overcome; therefore the system jumps to the
final running state for lower external forces. Numerical results
plotted in Fig.~\ref{fig3} for the case $\theta=21/31$ scale with the
expression $F_r=F_r^0-\xi \sqrt{T}$, if we chose $F_r^0=0.37$ and
$\xi=0.44$. While such a law is valid for other concentrations as
well, the parameters significantly depend on the concentration
contrary to the law for $F_\eta$.

The behavior of the 1D version of the problem is almost identical to
the 2D case. The only difference that we have to notice is in the
transition to the final sliding state. The exact critical force
depends slightly on the external conditions, which means that the
transitions do not occur simultaneously in all the channels. A careful
examination of the behavior in different channels shows two
interesting results. First, when a channel has jumped to the sliding
state, it enhances the probability for the neighboring channels to jump
due to interatomic interactions. This first channel corresponds
therefore to a nucleation event. Indeed, the width of the ``river'' of
neighboring moving channels grows faster than the number of
independent rivers.  The inset of Fig.~\ref{fig3} presents the 
numerical results
which scale very well with an exponential law $\exp \left[
\left(F-F_0\right)/\Delta F \right]$ with $F_0=0.276$ and $ \Delta
F=0.002$. Of course, the value of $\Delta F$ attests that it is a very
thin effect, but this phenomenon reveals an enhanced transition to
sliding state due to cooperative effects in the second dimension. On
the contrary, the transition of a channel to the locked state is
almost independent of neighboring channels.

In conclusion, a driven system of interacting particles in a 2-D
anisotropic external potential exhibits several dynamical transitions
succesfully explained by a hierarchy of depinnings. This behavior,
which is much richer than the behavior of a single Brownian particle
illustrates the interesting properties of complex systems. The analysis
of the transitions can be put in the framework of a single particle
theory by introducing collective excitations which have particle-like
properties although the strong discreteness of the system does not
allow us to consider these excitations as solitons. The complex
behavior of the multi-particle system is however not lost because,
after the transition to the full running state, the collective
organization is completely destroyed so that the backward transition
does not look like the upward sequence of transitions. This explains
why the complex system can maintain an hysteresis at non-zero
temperature contrary to the single Brownian particle.  This result
which emphasizes the role of hysteresis at microscopic scale should be
related to the main role of hysteresis found in solid friction~\cite
{nozieres}. Work along this line is in progress.

\begin{figure}
\caption{Mobility versus force for a concentration $\theta=21/41$
($N=105$, $M=205$). The solid curve
corresponds to increasing force and
the dashed curve, to decreasing force.}
\label{fig1}
\end{figure}

\begin{figure}
\caption{Atomic trajectories in a given channel for the 2D
system with $\theta=34/47$ at $F=0.24$.
The black diamonds correspond to the trajectory of one atom.}
\label{fig2}
\end{figure}

\begin{figure}
\caption{The diamonds (triangles)
correspond to the position of the transition to the running
(locked) state
for different temperatures in the case  $\theta={21/31}$,
while the squares corresponds to the transition to the locked state
for $\theta={21/41}$. The threshold was chosen to be $B=0.9B_f$.
The solid and dashed curves correspond to the phenomenological approach
discussed in the text.
The inset shows the width of the ``river'' (the moving
neighboring channels) in the 2D case as a function of the driving force
and the solid line corresponds to the approximate expression.}
\label{fig3}
\end{figure}

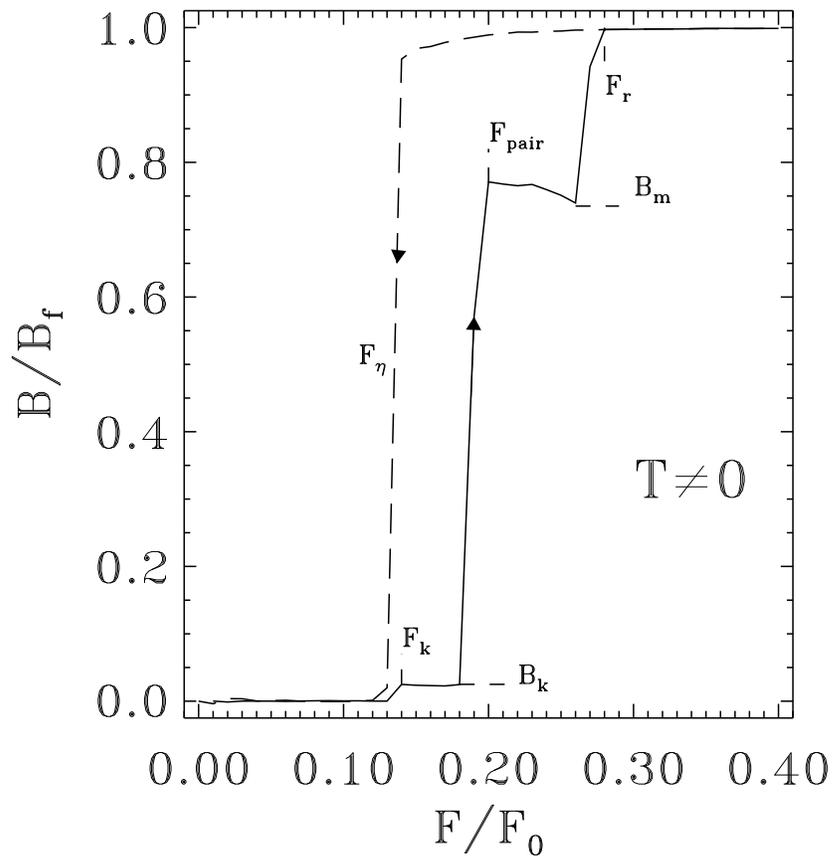
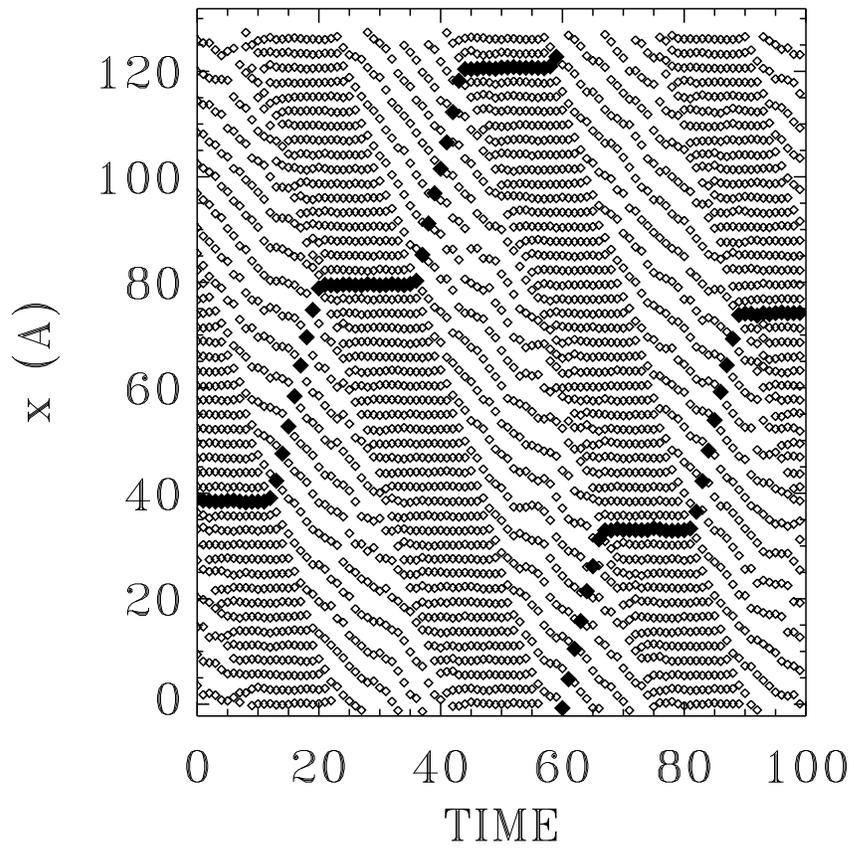
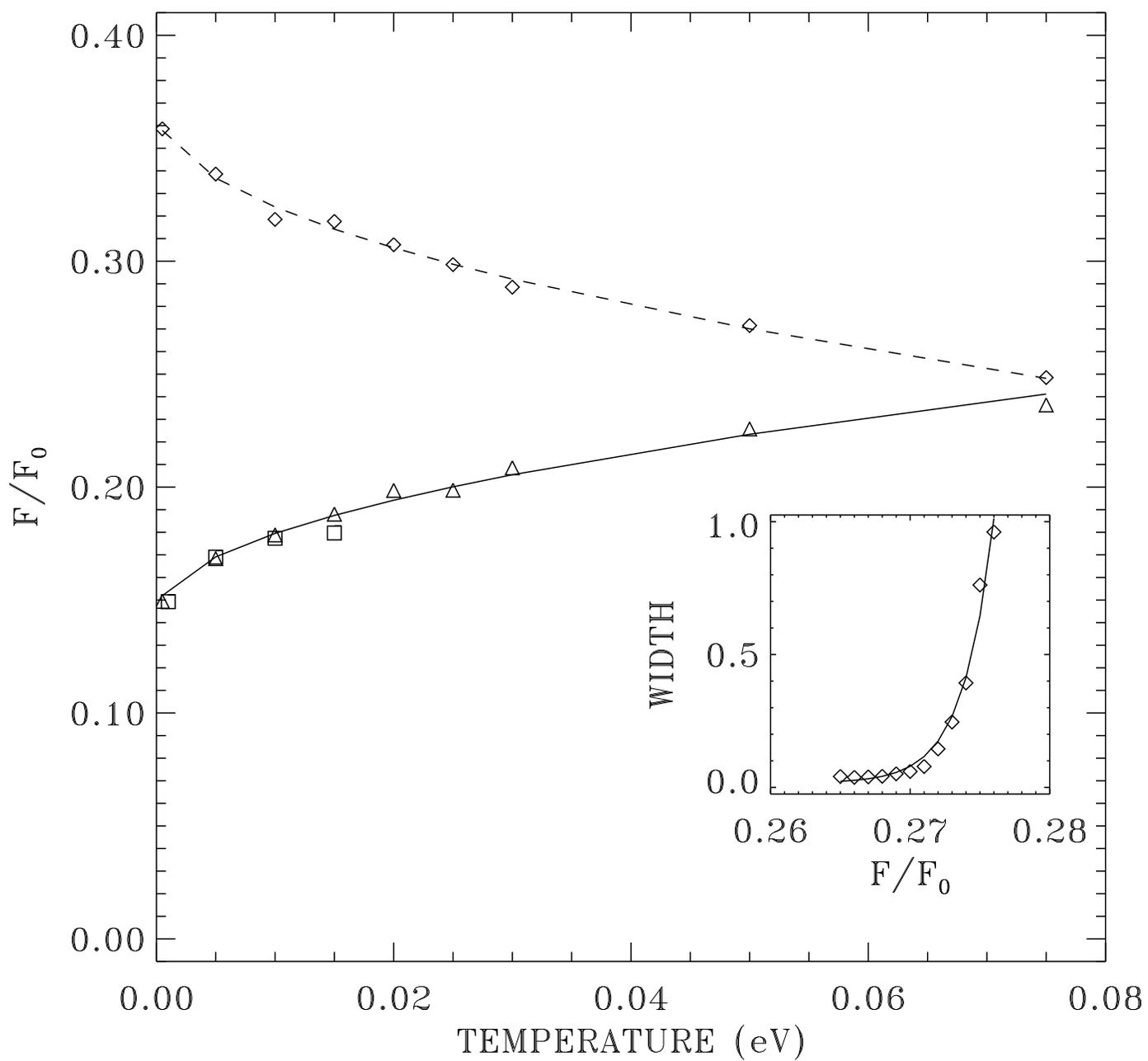
\end{document}